\documentclass{elsart}

\usepackage{graphics}
\usepackage{graphicx}
\usepackage{epsfig}

\usepackage{amssymb}
\usepackage{amsmath}

\begin{document}

\begin{frontmatter}



\title{High-energy neutrino emission from Gamma-Ray Bursts}

\author{F. De Paolis, }   
\author{G. Ingrosso, }    
\author{D. Orlando and}
\author{L. Perrone\corauthref{cor1}}

\address{Dipartimento di Fisica, Universit\`a di Lecce and INFN,
     Sezione di Lecce, 
Via Arnesano, CP 193, I-73100 Lecce, Italy}

\corauth[cor1]{ {\em Corresponding author} 
 - tel +390832-320459 - fax +390832-320505
}
\ead{lorenzo.perrone@le.infn.it}

\begin{abstract}

Gamma-Ray Bursts (GRBs) are expected to efficiently accelerate  
protons up to relativistic energies. 
High-energy photons can originate from decay of neutral pions
 produced by the interaction of 
 these protons with the medium surrounding the source. 
 In the same hadronic chain, high energy neutrinos are expected 
  to be produced  
 from decay of charged pions, as well as from 
  decay of generated muons.
  Neutrinos can travel cosmological distances being much less absorbed 
   than photons, thus providing a powerful tool for the investigation  
 of cosmic sources. 
In the frame of a recently proposed hadronic model for the emission 
 of high-energy $\gamma$-rays,     
  we estimate the neutrino emission from GRBs 
and calculate the neutrino 
fluxes at Earth as a function of the model parameters. The detectability of 
the expected signal by 
 current and future experiments is also extensively discussed. 
\end{abstract}


\end{frontmatter}


\section{Introduction}

Nowadays, it is widely recognized that GRBs have cosmological origin, 
as suggested by Beppo-SAX observations \cite{costa97} 
of counterparts in the X-ray band of some GRBs.
Thanks to these observations, it has been also possible to determine 
with great accuracy the GRB position and consequently, through  
 pointing 
of optical and radio telescopes, to evaluate the distance of some GRBs. 
The derived amount of energy radiated by assuming an isotropic explosion 
can result in some cases of the order of 
  ${\mathcal E}^{iso}_{\gamma} \sim 10^{53} - 10^{54}$ 
 erg in the soft $\gamma$-ray (50 - 300 keV) energy band,   
showing that in a GRB event up to one solar mass can be radiated in 
form of soft $\gamma$-rays.
Note that if GRBs are beamed sources as indicated by afterglow 
observations \cite{huang}, the  emitted energy would be 
decreased by the factor $\Delta \Omega / 4 \pi$, where $\Delta \Omega$ is 
the beaming angle. 

According to the fireball model for GRBs \cite{rees92}, 
relativistic 
electrons  are accelerated at 
shock fronts occurring near the interface of the expanding relativistic 
shells (external shock models), or at shocks forming within the unsteady 
outflow itself (internal shock models). 
Soft $\gamma$-rays are then produced   
through leptonic processes, such as synchrotron 
or Inverse Compton emission.
 
Many GRBs have been recognized as  
 high-energy
$\gamma$-rays emitters by the EGRET experiment which
detected $\gamma$-rays in the energy band between 30 MeV and  20 GeV 
from a high percentage of the bright bursts found by BATSE
 \cite{schneid1,schneid2,hurley}.
These observations clearly show that the GRB phenomenon 
is not exclusively the domain of soft-energy leptonic processes but 
may include high-energy hadronic mechanisms.\\
It has been suggested (see e.g.~\cite{galeotti})
 that highly relativistic protons accelerated at shock front  
  can produce quite efficiently 
very high-energy photons and neutrinos through $p\gamma$ interactions
within the dense photon field arising from the GRB \cite{wb97},
or by mutual $pp$ interactions in the ejecta \cite{px94}, 
or by $pn$ inelastic collisions with 
neutrons, if present in the original explosion \cite{bm00}.\\
The emission of high-energy neutrinos and $\gamma$-rays  
 has been also investigated within the
 so-called cannonball scenario, by which jets of highly relativistic 
  balls emitted in supernovae explosions collide 
  with the surrounding shell producing quasi thermal radiation
   (due to the collision heating), and  high energy 
  $\gamma$-rays and neutrinos (through
 $pp$ interaction in the shell) \cite{rujula}  
 
 Very recently, a model
  for the production of  high-energy $\gamma$-rays in GRBs  
 with $E_{\gamma}>1$ GeV has been proposed~\cite{noi}:
 high-energy protons emerging from the GRB source  
may subsequently interact with external nucleons, giving rise 
to neutral pions, which in turn promptly decay into high-energy photons. 
A key requirement of the model is the existence of 
a dense enough cloud (as a target for $pN$ interactions), 
which surrounds the GRB source, or, alternatively, situated along 
the line of sigh to the observer and near enough the source.
A delay in the arrival times of the 
{high-energy $\gamma$-rays} with respect to the {soft $\gamma$-rays} 
and a temporal spread of the signal at Earth is predicted 
  as a consequence of the proton 
deflection due to the presence of magnetic fields in the region surrounding 
the GRB source. 
Clearly, in addition to neutral pion production, 
inelastic $pN$ collisions also produce
charged pions, kaons and small quantities of other mesons.
If their main free path for interaction 
$\lambda_{i}= (n_N \sigma_{\pi N} )^{-1}$, 
is much larger than their decay path $\lambda_d = \Gamma \tau c$ 
(with $ \Gamma = E_{\pi} / m_{\pi} c^2$), they decay into 
charged leptons and neutrinos \cite{cr}
\footnote{
Note that even in the case of a cloud with density as high as
$n_N \simeq 10^{11}$ cm$^{-3}$, the condition 
$\lambda_d / \lambda_i <<1 $ is well verified, 
so that all the produced secondary unstable mesons 
decay before interacting with the cloud itself.}.
Here $n_N$ is the nucleon number density, $\sigma_{\pi N}$ is the 
 pion-nucleon cross section and $\tau$ is the pion proper mean life.   
The description of this model is given with some detail 
 in section 2.  In section 3,  
 we estimate the neutrino emission 
from GRBs as expected within this frame and compare 
 our results with  
the theoretical predictions by other models for the 
   production of neutrinos from astrophysical sources.       
   In section 4 we finally discuss the detectability of the expected signal 
 in terms of upward-going  neutrino-induced muons
 by taking into account 
 the experimental  
  constrains given by underground and underwater/ice detectors.

\section {The Model}

According to~\cite{noi}, we assume that the GRB source
emits an amount of energy ${\mathcal E}_p \simeq 10^{54} \Delta \Omega$ erg
in form of relativistic protons~\cite{totani},
released at a distance $r_0 \simeq 10^{16}$ cm from the central source, 
during a time $\Delta t\simeq 1$ s.

Moreover, we require that there exists a dense enough cloud near/around the GRB 
source (with $n_N \sim 10^{9}-10^{11}$ cm$^{-3}$)  
so that accelerated protons subsequently interact with the nucleons  
in the cloud ($\sigma_{pN} \sim 3 \times 10^{-26}$ cm$^{2}$),  
giving rise to neutral and charged pions and ultimately 
to high-energy photons and neutrinos
\footnote
{The average length travelled by protons in the surrounding 
medium results to be $ \Delta R =(\sigma_{pN} n_N)^{-1} \simeq
      10^{-3}~{\rm pc}~(10^{10}~{\rm cm}^{-3}/n_N)$. Correspondingly,
the mass enclosed will be $M_{\rm cl}\sim 0.4~M_{\odot} (r_0/10^{15}~
{\rm cm})^2$ in a shell-like cloud geometry around the GRB and 
$M_{\rm cl}\sim 0.2~M_{\odot} (10^{10}~{\rm cm}^{-3}/n_N)^2$ in the case of 
an intervening cloud near the GRB source.}.

The initial proton flux $J_p(E_p,r)$ is taken to be a function of both the
proton energy $E_p$ and the radial coordinate $r$ as in the following: 
\begin{equation}
\begin{array}{ll}
J_{p}(E_p,r) = C~(E_p/{\rm GeV})^{-a_p} f(r) 
~~~~~~~~~[{\rm protons~cm^{-2}~s^{-1}~sr^{-1}~GeV^{-1}]}~.
\end{array}
\label{eqno:pflux}
\end{equation}
Here, the radial profile, for $r \geq r_0$, has the form
\begin{equation}
f(r) = 
\left\{\begin{array}{l}
\displaystyle{\left(\frac{r_0}{r}\right)^2}
~\exp[-\sigma_{pN}~ n_{N}~ l(E_p,r)]
          ~~~~~{\rm if} ~~l<R_L(E_p) 
\nonumber \\
0~~~~~~~~~~~~~~~~~~~~~~~~~~~~~~~~~~~~~~~~~~{\rm if} ~~l>R_L(E_p)~,
\nonumber 
\end{array}
\right.
\label{eqno:fr}
\end{equation}
where $R_L(E_p) = E_p/(eB)$ is the proton Larmor radius, 
 $l(E_p,r) \equiv \alpha(E_p,r) R_L(E_p)$ 
is the path-length travelled 
by protons starting at $r_0$ up to their first $pN$ 
interaction occurring at distance $r$ from the source and  
$\alpha(E_p,r)$ is the proton deflection angle due to 
 the intervening magnetic field B (tipically $B \simeq 1 \mu$G)
 given by: 
\begin{equation}
\alpha(E_p,r) \simeq \arcsin \left(\frac{r-r_0}{R_L(E_p)} \right)~,
\end{equation}

As far as the value of the energy spectral index $a_p$ is concerned, 
since protons are accelerated by shock waves moving with ultra-relativistic 
velocities (corresponding to GRB Lorentz factor 
$\Gamma \simeq 300$,  see e. g. \cite{piran}
and diffusive processes are negligible during the time before $pN$ 
interactions, we adopt the relevant value  
$a_p \simeq  2.2$  \cite{bedn,vietri2}.

Finally, the constant $C$ in eq. (\ref{eqno:pflux}), 
 results from the normalization condition 
\begin{equation}
\Delta \Omega ~ r_0^2
\int_{E_0}^{+\infty} J_p(E_p,r_0)~ E_p ~ dE_p =
 \frac{ {\mathcal E}_p}{\Delta t}~,
\end{equation}
where $E_{0} = \Gamma m_{p} c^2$ is the minimum proton energy.

Following Dermer \cite{dermer}, the production spectrum of 
{high-energy $\gamma$-rays} resulting from 
$pN$ collisions, via $\pi^0 \rightarrow \gamma \gamma$ decay,
is given by the source function
\begin{equation}
\begin{array}{ll}
q_{\gamma}(E_{\gamma},r)=
16 \pi^2 n_{N} 
\displaystyle{\int_{E_{\pi}^{min}}^{+\infty}} dE_{\pi}
\displaystyle{\int_{E_{0}}^{+\infty}} dE_p~ J_p(E_p,r) \times \\ \\
\displaystyle{\int_{cos\theta_{max}}^{1}} dcos\theta~ E^{*} 
\displaystyle{ \frac{d^3 \sigma^{*}}{dp^{*3}} }
~~~~~~~~~[{\rm \gamma~cm^{-3}~s^{-1}~GeV^{-1}}]~,
\label{eqno:qgamma}
\end{array}
\end{equation}
where $\theta$ is the angle between the line of sight and the proton direction,
asterisks refer to CM frame and $E^{*} d^3\sigma^*/dp^{*3}$ is the Lorentz 
invariant cross-section for $\pi^0$ production in high-energy 
($E_p \lesssim 10^3$ GeV) $pN$ collisions 
\footnote{
Mori \cite{mori} had verified that extrapolation up to 
$E_p \simeq 10^6$ GeV of the invariant cross-section used by 
Dermer \cite{dermer} is in reasonable agreement with Monte Carlo 
simulations by several $pN$ event generators.}.

Numerical values of $q_{\gamma}(E_{\gamma},r)$
depend on the main model parameters, namely the GRB energetics ${\mathcal E}_p$, 
the nucleon number density $n_N$ and the magnetic field strength $B$
(for more details see \cite{noi}).

An important effect of the presence of magnetic fields
is a temporal spread of the emitted {high-energy $\gamma$-ray}
signal on Earth. To better analyze this point, we can consider the proton 
injection as instantaneous, since the acceleration time
$\Delta t \simeq 1$ s is much smaller 
than the average proton interaction time 
$(\sigma_{pN} n_N c)^{-1} \simeq 10^5~{\rm s}~(10^{10}~{\rm cm}^{-3}/n_N)$.
Accordingly, denoting by $t_0$ the instant at which photons arrive 
on Earth in the case $B=0$ (no proton deflection), 
the delay $t_i-t_0$ can be expressed as a function of the 
radial coordinate $r_i$ at which $pN$ interactions occur 
and the proton Larmor radius $R_L(E_p)$:
\begin{equation}
\begin{array}{ll}
t_i-t_0 
= \displaystyle{ \frac{R_L(E_p)}{c}
\left(\arcsin \frac{r_i-r_0}{R_L(E_p)} - \frac{r_i-r_0}{R_L(E_p)}\right)}~.
\end{array}
\label{eqno:delay}
\end{equation}
Vice versa, one can say that photons which arrive on Earth within  
the time window ($t_0$,$t_i$) are
produced by protons of energy $E_p$ only if these protons
have interacted with nucleons in the shell between $r_0$ and $r_i(E_p)$.

Correspondingly, the source spectrum 
$Q_{\gamma}(E_{\gamma}; t_0,t_i)$ of $\gamma$-rays 
produced in a GRB event within the temporal window $t_i -t_0$  
is given by  
\begin{equation}
\begin{array}{lll}
Q_{\gamma}(E_{\gamma}; t_0,t_i) =
16 \pi^2 n_N ~\Delta t 
\displaystyle{\int_{E_{\pi}^{min}}^{+\infty}} dE_{\pi}  
\displaystyle{\int_{E_0}^{+\infty}} dE_p ~ \times \\ \\
\displaystyle{\int_{r_0}^{r_i(E_p)}}~ dr J_p(E_p,r)~\exp[-\sigma_{pN} n_{N} l(E_p,r)] 
~\times \\ \\
\displaystyle{\int_{cos\theta_{max}}^{1}} dcos\theta~ E^{*} 
\displaystyle{\frac{d^3 \sigma^{*}}{dp^{*3}}}
~~~~~~~~~~~~~~~~~~~~~~~~{\rm [\gamma~GeV^{-1}]}~,
\end{array}
\label{eqno:lum}
\end{equation}
with the proton injection time $\Delta t\simeq 1$ s.

\begin{figure}[htbp]
\begin{center}
\epsfig{file=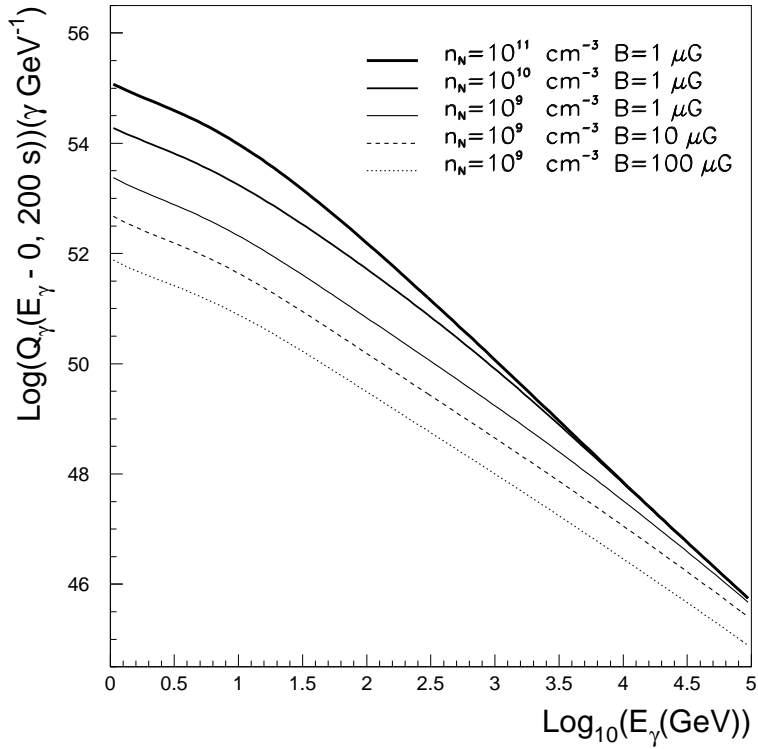,height=11cm,width=11cm}
\caption{
  The source $\gamma$-ray spectrum $Q_{\gamma}(E_{\gamma}; 0,~200 \,{\rm s})$ 
  in the first 200 s of the signal life 
  for different values of the cloud number density  
  $n_N=10^9-10^{11}$ cm$^{-3}$ and of the magnetic field 
  strength $B=10^{-6}-10^{-4}$ G. 
  The GRB total energetics is assumed to be 10$^{54}$ erg/sr.
}
\label{Figure1}
\end{center}
\end{figure}

In Fig. \ref{Figure1}, assuming different values for the model parameters $n_N$ and $B$ 
in the range $10^9 -10^{11}$ cm$^{-3}$ and $10^{-6}-10^{-4}$ G, 
respectively, we show $Q_{\gamma}(E_{\gamma}~;~0,~200~{\rm s})$ 
in the first 200 s, starting from the inset of the signal $t_0 =0$.
From Fig. 1, one can see that for a given photon energy $E_{\gamma}$
the differential $\gamma$-ray production 
increases with increasing $n_N$ and decreasing $B$.
The effect is particularly important at low $E_{\gamma}$ values.
In fact, low-energy photons are mainly generated by low-energy protons
with $E_p$ roughly one order of magnitude above $E_{\gamma}$: they
are more easily deflected by the intervening magnetic fields 
giving rise to photons that cannot reach the observer
since $\alpha(E_p,r) > \pi/2$. 
However, our numerical results show that in a large range of values for 
$n_N$ and $B$, the decrease of $Q_{\gamma}(E_{\gamma}; t_0,t_i)$ 
for increasing values of $B$ can be 
counterbalanced by increasing $n_N$. Moreover,
for each assumed value of $B$, there exists a limiting density 
($n_N \simeq 10^{11}$ cm$^{-3}$ in the case $B \simeq 1~\mu$G)
above which most of the protons interact before suffer substantial
deflection.

Note that in deriving eq. (\ref{eqno:lum}) we have neglected 
re-absorption of the produced $\gamma$-rays in the cloud itself
\footnote{
High-energy $\gamma$-rays may in 
principle be re-absorbed through the interactions with the atomic 
 hydrogen,
provided that the path they have to travel inside the cloud 
(after being produced) is long enough.
However, clouds nearby the GRB source are expected to present 
a high ionization degree, so that the relevant absorption processes are 
$\gamma e$ and $\gamma p$ interactions, which imply
a large $\gamma$ free-path before re-absorption.}.
Clearly, re-absorption of neutrinos within the cloud itself
can be completely neglected.

\section {High-energy neutrino emission}

It is well known that, for a proton power law spectrum
$J_p(E_p) \propto E_p^{-a_p}$ and inclusive cross sections 
for meson production obeying Feynman scaling,
the produced spectrum of $\gamma$-rays, $\nu_e$'s, $\nu_\mu$'s 
and $\nu_\tau$'s are all proportional at high energies 
(see e. g. \cite{gaisserbook})
Following the calculation developed by Dar~\cite{dar} and by 
Lipari~  \cite{lipari}, we assume in particular that muon 
neutrino and gamma ray 
  spectra by $pN$ interactions  are related 
through $Q_{\nu_{\mu}} \sim K Q_{\gamma}$ with $K \sim 0.7$. 
From eq. (\ref{eqno:lum}),   
we can then estimate the differential neutrino fluences on Earth
within the time interval $t_i -t_0$
\begin{equation}
\begin{array}{ll}
F_{\nu}(E_{\nu};t_0,t_i) = 
\displaystyle{ \frac{ Q_{\nu}[E_{\nu}(1+z);t_0,t_i]~(1+z)}
{4 \pi D_L^2(z)} }
~~~~~~~~~~~~{\rm [\nu~cm^{-2}~GeV^{-1}]}~,
\end{array}
\label{eqno:flux}
\end{equation}
where $z$ is the GRB redshift and
\begin{equation}
D_L(z) = \frac{c}{H_0 q_0^2} 
         \left(z q_0 + (q_0-1)(-1+\sqrt{2 q_0 z +1})\right)
\label{eqno:DL}
\end{equation}
is the GRB luminosity distance, with $H_0 \simeq 70$ km s$^{-1}$ Mpc$^{-1}$ 
and $q_0 \simeq 0.15$. 
In Fig. 2, assuming  $z=1$ and ${\mathcal E}_p = 10^{54}$ erg/sr, 
the neutrino fluence $F_{\nu}(E_{\nu}$; 0, 200 s) 
in the initial 200 s is given as a function of neutrino energy 
for selected values of 
$n_N$ and $B$ parameters. A comparison with Fig. 1
shows that, 
apart the energy $z$ shift,
the neutrino fluence 
follows, as a function of model parameters, 
 the same behaviour of the source $\gamma$-ray spectrum.  
\begin{figure}[htbp]
\begin{center}
\epsfig{file=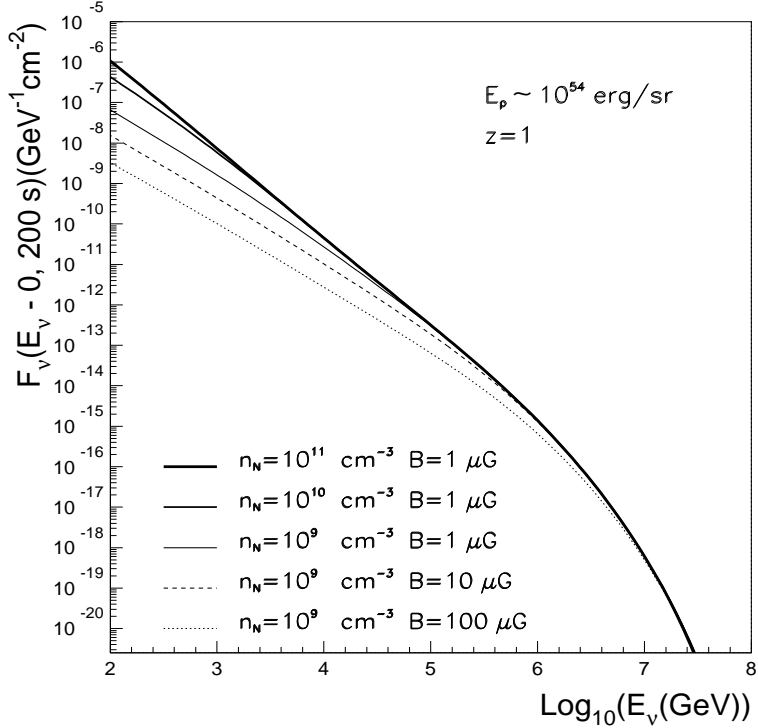,height=10.5cm,width=11cm}
\caption{
  Expected $\nu_{\mu}+\bar{\nu}_{\mu}$ 
  neutrino fluences $F_{\nu}(E_{\nu};$ 0, 200 s) 
  in the first 200 s of the signal life 
  for different values of the cloud number density  
  $n_N=10^9-10^{11}$ cm$^{-3}$ and of the magnetic field 
  strength $B=10^{-6}-10^{-4}$ G. 
  The GRB total energetics and redshift are assumed to be 10$^{54}$ erg/sr
  and $z=1$, respectively.
  }
\label{Fluenze}
\end{center}
\end{figure}

Starting from the neutrino fluence $F_{\nu}(E_{\nu};t_0,t_i)$
during the time interval $t_i - t_0$, 
one can estimate an average neutrino flux on Earth from a single 
 source as follows: 
\begin{equation}
< \Phi_{\nu}(E_{\nu}) >~=~\frac {F_{\nu}(E_{\nu};t_0,t_i)}{t_i-t_0}~.
\end{equation}

Otherwise, it is also possible that neutrinos from unresolved GRBs sources  
give a contribution to the diffuse high energy neutrino background. 
In this respect, due to the widely accepted cosmological origin of 
 GRBs, we adopted, as recently done in 
  \cite{stecker2000}, the cosmological evolution 
scheme developed by Schmidt \cite{schmidt}.   
In particular, we assume that
the rate of GRB events, per unit comoving volume and per unit 
comoving time, is given by 
\begin{equation}
\dot{n}(z) = 
\left\{\begin{array}{l}
\dot{n}_0 (1+z)^p   ~~~~~~~~~{\rm for}~  0 < z < z_p
\nonumber \\
\dot{n}_0 (1+z_p)^p  ~~~~~~~~{\rm for}~    z > z_p
\nonumber \end{array}
\right.
\end{equation}
with $z_p=1$ and $p=3.32$. The parameter $\dot{n}_0$
is then determined so that the observed (in the soft $\gamma$-ray band)
total GRB event rate 
\begin{equation}
\dot{N}_{GRB} = \int_0^{z_{max}} \frac{\dot{n}(z)}{1+z} \frac{dV}{dz}~dz 
\label{eqno:rate}
\end{equation}
reads $\dot{N}_{GRB}\simeq 10^3$ yr$^{-1}$. 
Here, $dV = 4 \pi D_M^2 dD_M$ is the comoving volume element,
$D_M = D_L / (1+z)$ is the proper motion distance, related 
to the luminosity distance $D_L$ given in eq. (\ref{eqno:DL}), $z_{max}$
 is taken equal 5.   
Moreover, in eq. (\ref{eqno:rate}) we also take into account that 
the time interval between  bursts is stretched by the factor 
$(1+z)$.

For simplicity, we assume that GRBs are standard candles,
each source with ${\mathcal E}_p = 10^{54}$ erg/sr, $B = 1 \mu $G and
$n_N = 10^9$ cm$^{-3}$.
Correspondingly, the diffuse neutrino flux expected on Earth results to be
\begin{equation}
\Phi_{\nu}(E_{\nu}) = 
\int_0^{z_{max}} \frac{ Q_{\nu}[E_{\nu}(1+z); 0, +\infty] }{4 \pi D_L^2(z)} ~
\frac{\dot{n}(z)}{1+z} \frac{dV}{dz}~dz
\end{equation}
\begin{figure}[htb]
\begin{center}
\epsfig{file=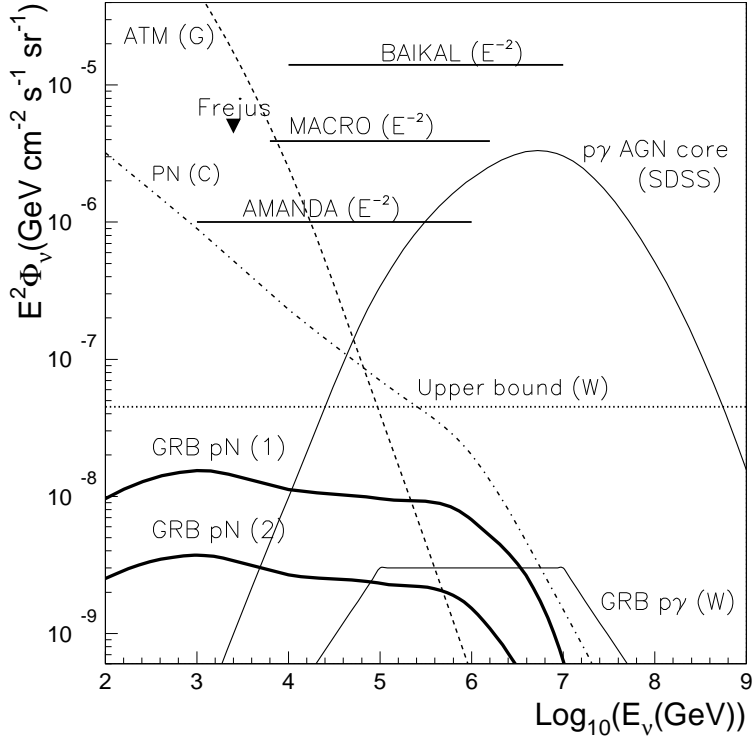,height=11cm,width=11cm}
\caption{Diffuse fluxes of $\nu_{\mu}+\bar{\nu}_{\mu}$ from 
this model assuming no evolution [GRB pN (1)] and 
 Schmidt's evolution [GRB pN (2)], respectively. 
The predictions according to other models of 
AGNs and GRBs are also shown: SDSS~\cite{Stecker}, 
W~\cite{wb98}.
  The dashed curve refers to the angle average atmospheric
neutrino flux~\cite{gaisser}; 
  the dash-dotted line refers to the flux of prompt neutrinos 
   according to  one of the highest predictions given in
\cite{costa};   the dotted line is the theoretical upper bound to neutrino
 flux from astrophysical sources as calculated 
in  \cite{wb98}. Some current experimental upper limits are also
shown: AMANDA \cite{now2000}, MACRO \cite{miolimite}, Frejus \cite{Frejus}
and BAIKAL~\cite{baikal}.}
\label{fondo}
\end{center}
\end{figure}
In Fig.~\ref{fondo}, $\Phi_{\nu}(E_{\nu})$ is given as a function of
 $E_{\nu}$.
For comparison we show the diffuse neutrino flux from GRB events    
(via $p\gamma$ interactions) predicted by the Waxman \& Bahcall model 
\cite{wb97} as well as the upper bound implied by high energy 
cosmic ray observations    
estimated by the same authors~\cite{wb98}. 
Moreover, in Fig.~\ref{fondo}, we give the 
  atmospheric neutrino flux averaged on the zenith angle as calculated 
by Gaisser et al. \cite{gaisser}. Actually, 
at energy above 10 TeV, 
the isotropic contribution of neutrinos produced in 
the atmosphere by  
the semileptonic decays of charmed 
hadrons (prompt neutrinos),
starts to dominate  
over the conventional pion and 
kaon decay induced atmospheric neutrino flux.   
In Fig.~\ref{fondo} we report the contribution  
of prompt neutrinos as calculated 
 in \cite{costa} 
within the frame of the 
Recombination Quark Parton Model.
It should be remarked that 
there is some theoretical uncertainty on 
 the calculation of prompt neutrinos flux, as well 
 as on the energy at which this contribution 
 crosses over the conventional atmospheric flux. 
In Fig.~\ref{fondo}, we considered   
 one of the highest prediction among those 
 given in \cite{costa}. 
 Finally, the upper limits  
from current experiments
(Frejus \cite{Frejus}, AMANDA \cite{amanda} and BAIKAL \cite{baikal})
are shown, derived by assuming 
a power law spectrum ($\propto E_{\nu}^{-2}$)  
an initial neutrino flux.

\section{Neutrino signal detectability}

\subsection{Neutrino propagation} 
\begin{figure}[htb]
\begin{center}
\epsfig{figure=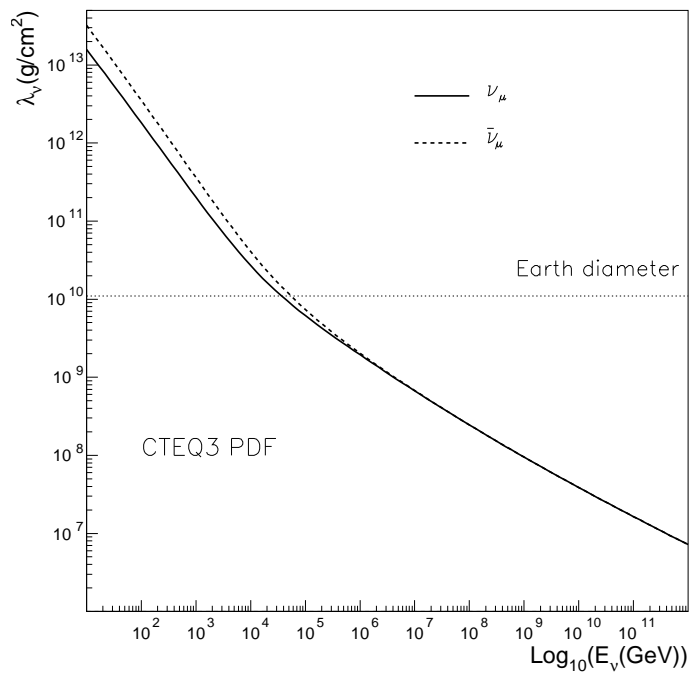,height=10cm,width=10cm}
\caption{Neutrino interaction length 
   for deep inelastic scattering off an isoscalar 
   nucleon $N=(n+p)/2$, for      
   $\nu_{\mu}$ (solid line) and  
   $\bar{\nu}_{\mu}$ (dashed line).  
    The calculation has been done for the CTEQ3 
   set of parton density functions~\cite{cteq}. 
     The dotted line gives the Earth diameter.}
\label{lamneu}
\end{center}
\end{figure}

High-energy neutrinos from astrophysical 
sources can be detected by collecting
the upward-going  neutrino-induced muons produced 
through charge-current interactions     
with the matter surrounding the detector. 
Based upon this detection principle, current underground 
and underwater/ice experiments   
 are surveying the sky in order to search  for 
 neutrino sources. 
In the frame 
of the model described above, 
we have investigated by numerical 
calculation the detectability of the signal 
 expected either from a single point-like GRB, or from  
 the diffuse flux due to unresolved GRBs.      
Neutrino propagation   
 through the Earth has been developed following the approach  
  discussed in~\cite{naumov}, where they give an  
   analytical method for the precise calculation of the spectrum 
  of high energy neutrinos propagating through a dense medium of any
  thickness, for initial spectrum  and for cross
   sections of any form.\\ 
 As a preliminary observation, it should be noticed
         that the results of   
   atmospheric neutrino experiments~\cite{SK,Ambrosio98} 
    suggest that muon neutrinos 
    can oscillate into 
   tau neutrinos, thus affecting       
   the expected muon event rate on the detector.   
    This effect has been extensively discussed e.g. in \cite{quiggtau}.
    Here, flavour oscillation will be not considered.\\
At energy above 1 GeV neutrino-matter interaction 
is dominated by the process of 
 {\it deep inelastic scattering} (DIS) on nucleons (charge and neutral 
  current processes), 
 since the contributions of both 
   elastic and     
 quasi-elastic 
 interactions 
  become negligible.  
The main difficulty for the evaluation of the 
deep inelastic neutrino-nucleon cross sections 
 arises from the fact that, in spite of the lack of experimental data, 
 at very high energy 
the details of nucleon structure  
 become important.\\
 Detectable muons are produced through charge current 
interaction as in eq.~(\ref{cc}). Neutral current interaction 
 described by eq.~(\ref{nc}) causes instead 
 a modulation in the spectrum of the interacting neutrinos.    
 \begin{equation}
\nu_{\mu} (\bar{\nu}_{\mu}) + N \rightarrow \mu^{-} (\mu^{+}) + X~, 
\label{cc}
\end{equation}
\begin{equation}
\nu_{\mu} (\bar{\nu}_{\mu}) + N \rightarrow \nu_{\mu} (\bar{\nu}_{\mu})
 + X'~.
\label{nc} 
\end{equation}
The total cross section for these processes can be written 
 in terms of differential cross sections 
as follows: 
\begin{equation}
\sigma_{tot}^{cc}(E_{\nu})=  \int_{0}^{1-\frac{m_{\mu}}{E_{\nu}}}
              \frac{d\sigma^{cc}}{dy} \,(E_{\nu},y) \, dy ~,
\end{equation}
\begin{equation}
\sigma_{tot}^{nc}(E_{\nu})=  \int_{0}^{1}
             \frac{d\sigma^{nc}}{dy} \,(E_{\nu},y) \, dy~, 
\end{equation}
where  $E_{\nu}$ is the energy of the incoming neutrino, $m_{\mu}$ is the 
   muon mass and 
 $y$ is defined by $E_{l}/E_{\nu}$ with $E_{l}=E_{\nu}-E_{\mu}$
  for charge current interaction ($E_{\mu}$ is the energy of the outcoming muon)
 and $E_{l}=E_{\nu}-E_{\nu}'$ for neutral current interaction 
   ($E_{\nu}'$ 
  is the energy of the outcoming neutrino).\\ 
Let $F_{\nu}(E_{\nu},x)$ be the differential energy 
spectrum of neutrinos at a column 
depth $x=\int_{0}^{L} \rho(L')dL'$, 
where $\rho(L)$ is the density of the medium at distance $L$ from the 
 boundary, measured along the neutrino beam path. The one-dimensional 
  transport equation can be expressed in the form~\cite{naumov}:
\begin{equation}
\frac{\partial\, F_{\nu}(E_{\nu},x)}{\partial x}=-\frac{F_{\nu}(E_{\nu},x)}
{\lambda_\nu (E_{\nu})}+
    \int_{0}^{1} \frac{N_{A}}{A_{N}} \,
             \frac{d\sigma^{nc}}{dy} 
	      F_{\nu}\left(\frac{E_{\nu}}{1-y},x\right) \,
\frac{dy}{1-y}~,
\label{gente}
\end{equation}
with the boundary condition $F_{\nu}(E_{\nu},0)=F_{\nu}^{0}(E_{\nu})$.
Here, $\lambda_\nu (E_{\nu})$ is the the neutrino interaction length 
 in g cm$^{-2}$
defined by: 
\begin{equation}
\lambda_\nu(E_{\nu})=\frac{1}{(N_{A}/A_{N}) \, (\sigma_{tot}^{cc}(E_{\nu})
 + \sigma_{tot}^{nc}(E_{\nu}))}~, 
\label{meanfree}
\end{equation}
where $N_{A}$ is the Avogadro number and $A_{N}=1$ g mol$^{-1}$.\\
The first term in the right side of eq.~(\ref{gente})  accounts  
for the absorption of neutrinos
     of energy $E_{\nu}$, due to charge and neutral current interactions;      
  the second term describes the additive contribution   
       of higher energy neutrinos      
             ``regenerating'' 
    through neutral current interaction at energy 
       $E_{\nu}$  . 
The solution of eq.~(\ref{gente}) is taken of the form  
\begin{equation}
F_{\nu}(E_{\nu},x) = F_{\nu}^{0}(E_{\nu})
   \exp\left[-\frac{x}{\Lambda_\nu(E_{\nu},x)}\right] , 
\label{Lambda}   
\end{equation}
where the function $\Lambda_\nu(E_{\nu},x)$ is    
 the     
effective absorbtion length containing the relevant information 
 on the modulation of the neutrino flux $F_{\nu}(E_{\nu},x)$
 along the propagation (see Ref.~\cite{naumov} for further details). 
 The differential $\nu_\mu N$ and $\overline{\nu}_\mu N$
cross sections have been calculated by using the approach in  
 Ref.~\cite{quiggcross} based on
the renormalization-group-improved parton model and on the last 
available experimental
information about the quark structure of the nucleon.\\
They are written in terms of the Bjorken scaling variables $y$ and 
 $\hat{x}=Q^2/2ME_{l}$, where $-Q^2$ is the invariant
  momentum transferred between the incoming neutrino and the outgoing 
   lepton (muon or neutrino, for charge and neutral current 
   interaction, respectively).    
     The detailed formulas used for the calculation are 
      given in \cite{quiggcross}.\\
Various versions of different sets of parton density functions
$q(\hat{x},Q^2)$ are collected in a large CERN software library
PDFLIB~\cite{PDFLIB97}; they can be simply accessed by setting few
parameters to choose the desired version. In this calculations, we  
selected  the third version of the
CTEQ collaboration model~\cite{cteq} for the  next-to-leading
order (NLO) quark distributions in the deep-inelastic scattering
factorization scheme. The CTEQ3 distributions are particularly
suitable for high energy calculations since the numerical evolution is provided
for very low $\hat{x}$. 
 Unfortunately, the uncertainty due
to this extrapolation may be rather large and is hard to estimate.
The $Q^2$ evolution is realized by the NLO
Gribov-Lipatov-Altarelli-Parisi equations from initial
$Q_0^2=2.56$ $GeV^2$.
(The new version of parton density functions by CTEQ 
group~\cite{cteq5} doesn't produce significant changes 
 for this calculation).\\ 
Fig.~\ref{lamneu} shows     
the neutrino interaction length $\lambda_\nu (E_{\nu})$
for deep inelastic scattering off an isoscalar 
   nucleon $N=(n+p)/2$ ($n \rightarrow$ neutron, 
   $p \rightarrow$ proton), for      
   $\nu_{\mu}$ (solid line) and  
   $\bar{\nu}_{\mu}$ (dashed line). The dotted line 
    placed at the level of the Earth diameter   
   demonstrates that 
 the Earth becomes opaque to neutrinos with energy exceeding 
   the TeV energy-range.    
%
\subsection{Muon propagation}
\begin{figure}[htb]
\begin{center}
\epsfig{figure=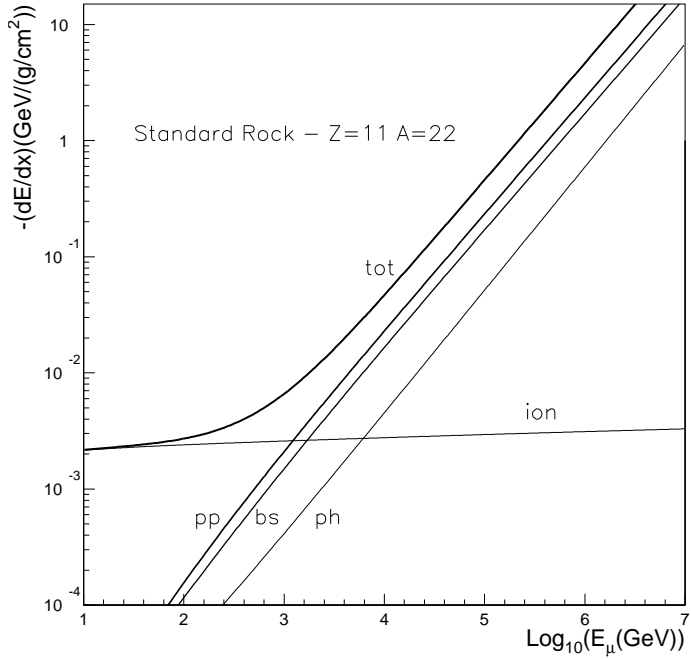,height=10cm,width=10cm}
\caption{Total average muon energy losses in standard rock 
 (thick line). The contributions of each electromagnetic 
  process have been also plotted,  
     $ion\rightarrow$ ionization, $pp \rightarrow$ $e^{+}e^{-}$ pair 
 production, $bs \rightarrow$ bremsstrahlung, $ph \rightarrow$ photonuclear
 interaction.}
\label{loss}
\end{center}
\end{figure}

High energy muons propagating through matter mainly interact   
by quasi-continuous (ionization) and discrete  
($e^{+}e^{-}$ pair production, bremsstrahlung and 
photonuclear interaction) electromagnetic scattering.    
Ionization dominates at energy lower than few hundreds of $GeV$, 
whereas above the $TeV$ region energy losses are mainly due to radiative
processes.\\
At this stage, muon propagation has been treated
 analytically,    
  according to the approach suggested in~\cite{naumovmuoni}.   
 Following~\cite{naumovmuoni}, the one-dimensional 
 transport equation can
  be written as:
\begin{equation}
\frac{\partial F_\mu(E_{\mu},x)}{\partial x}=
           \frac{\partial}{\partial E_{\mu}}(\beta(E_{\mu})F_\mu(E_{\mu},x))+
             \mathcal{G}(E_{\mu},x) , 
\label{mute}
\end{equation}
with the boundary condition 
\[ F_\mu(E_{\mu},0)=0 .\] 
Here, $F_\mu(E_{\mu},x)$
is the muon flux at energy $E_{\mu}$ and depth $x$, 
$\beta(E_{\mu}$) is the average 
total energy loss per unit length due to electromagnetic 
muon scattering (ionization plus radiative
processes) and    
$\mathcal{G}(E_{\mu},x)$  is the ``source'' function introduced to describe    
 the production of muons through neutrino charge current 
interaction. $\mathcal{G}(E_{\mu},x)$ is given by:
\begin{equation}
\mathcal{G}(E_{\mu},x)= \frac{N_{A}}{A_{N}} \,
 \int_{0}^{1} \frac{d\sigma^{cc}}{dy}
   \left( \frac{E_{\mu}}{1-y},y\right) \, F_\nu \left( \frac{E_{\mu}}
{1-y},x\right) \, \frac{dy}{1-y}. 
\label{source2}
\end{equation}
The solution of eq.~(\ref{mute}) can be written in the form:
\begin{equation}
\vspace{0.5cm}
F_\mu(E_{\mu},x)= \int_{E_{\mu}}^{\mathcal{E}(E_{\mu},x)}
       \frac{dE^{'}_{\mu}}{\beta(E_{\mu})}
       \, \mathcal{G}(E^{'}_{\mu},\,x-\mathcal{R}(E^{'}_{\mu},E_{\mu})), 
\label{solumu}
\end{equation}
where the function $\mathcal{R}(E^{'}_{\mu},E_{\mu})$ defined by     
\begin{equation}
\mathcal{R}(E^{'}_{\mu},E_{\mu})=\int_{E_{\mu}}^{E^{'}_{\mu}} \,
               \frac{dE^{''}_{\mu}}{\beta(E^{''}_{\mu})}
\label{rang}
\end{equation} 
 represents the mean depth 
 crossed by a muon of  
 initial energy $E^{'}_{\mu}$ and final energy $E_{\mu}$;   
$\mathcal{E}(E_{\mu},x)$ labels the initial energy needed by a muon 
 to reach the depth $x$ with final energy $E_{\mu}$. 
%
%
The calculation of muon energy losses is 
 crucial  
 to correctly evaluate the expected muon flux on the detector.    
 The muon energy losses due to radiative processes  
 (e.g. $e^{+}e^{-}$ pair production, bremsstrahlung and 
photonuclear interaction) 
  can be expressed in terms of differential cross section 
  for the specific muonic interaction process $k$ as follows: 
\begin{equation}
   -\Big\langle \frac{dE}{dx}\Big\rangle_{k}=\frac{N_{A}}{A} \, E_{\mu} \,
         \int_{v_{min}}^{v_{max}}\, v
 \frac{d\sigma^{k}}{dv}(v,E_{\mu}) \,dv , 
\label{losses}
\end{equation}
where $N_{A}$ and $A$ are Avogadro's number and the mass number respectively; 
$v$ is the fraction of initial energy $E_{\mu}$ lost by muon 
at the occurrence of the process $k$; 
$v_{min}$ and $v_{max}$ are the kinematical limits for the allowed values of
$v$; 
$x$ is the thickness of crossed matter, expressed in g cm$^{-2}$. 
The total average muon energy losses $\beta(E_{\mu})$ results 
 from the sum of the radiative interaction terms plus the 
  quasi-continuous contribution due to ionization. 
The explicit formulas used in this work   
 are given in~\cite{gmu}, accordingly to the  
 theoretical framework discussed in 
 the standard reference by Lohmann and Voss~\cite{Lohmann}. 
 Fig.~\ref{loss} shows the total average muon energy loss 
 (thick line) and the contributions from the single electromagnetic 
  process.

\subsection{Discussion and results} 

\begin{figure}[htb]
\begin{center}
\epsfig{file=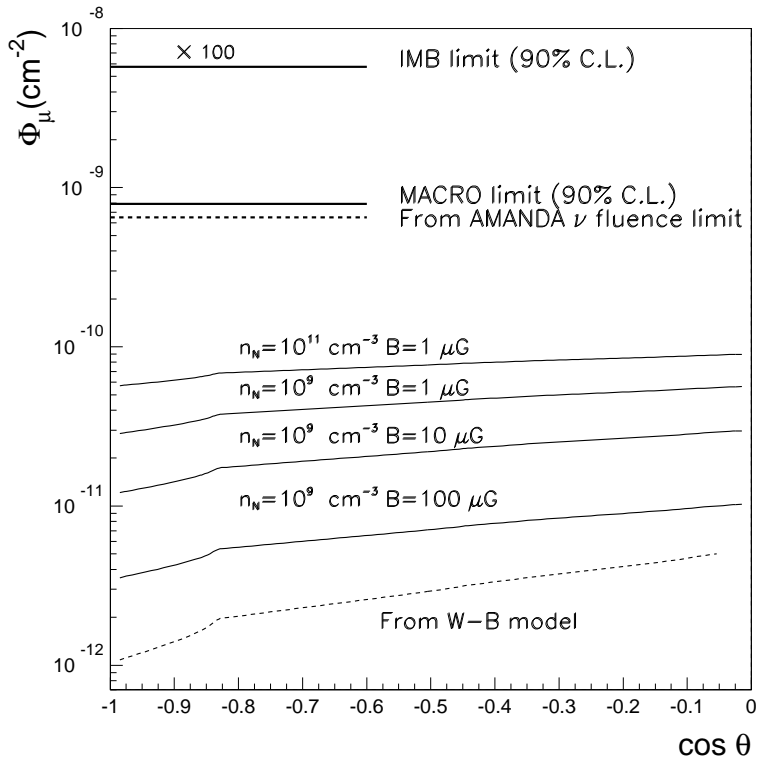,height=11cm,width=11cm}
\caption{The total number of events per cm$^{2}$ expected 
      from a single GRB      
     as a function of the cosine 
    of zenith angle $\theta$. The plot is shown for the 
     set of model parameters given in Fig.~\ref{Fluenze}.        
    The thin dashed line 
    is calculated starting from    
    the model by Waxman and Bahcall~\cite{wb98}.  
    Upper limits by  
  experiments MACRO~\cite{macro}, IMB~\cite{imb},
   and AMANDA~\cite{amanda} are also shown (see text for 
    comments).}
\label{limitiplot}
\end{center}
\end{figure}

In Fig. \ref{limitiplot}, assuming the same parameter values adopted   
 in Fig. \ref{Fluenze},  
 we give the expected number of 
 events per cm$^{2}$ from a single GRB source,  
  as a function of 
  the cosine of zenith angle $\theta$.

 The dependence on $\theta$ of the muon flux is a consequence of 
  neutrino absorption in Earth, which cannot be neglected at high energy. 
  Moreover, the suppression  
   at $\cos \theta \leq -0.8$
 reflects the hard discontinuity of Earth density profile  
  in correspondence of the massive nucleus. 
 In particular,  
 in Fig.~\ref{limitiplot} the calculated fluxes are compared with 
  upper limits by  
  experiments (MACRO~\cite{macro}, IMB~\cite{imb},
   and AMANDA~\cite{amanda}). 
 (Note that the IMB limit has been 
 lowered for convenience of presentation.
  Moreover,   
 the AMANDA limit, shown as a dashed line, has been estimated  
   starting from the declared upper limit 
   on neutrino flux).
   Finally, for comparison purposes,   
    in Fig.~\ref{limitiplot} we show the number of events calculated  
      by assuming as theoretical input the  
       well-known Waxman and Bahcall (\cite{wb97}) 
        model of neutrino production from GRBs.
The sensitivity to gamma ray bursts as point-like sources 
 is improved by the search for space and time correlation with optical  
 information. In this case the background is strongly suppressed.  

The same analytical procedure has been used in order to 
 calculate the expected number of events induced by our predicted 
 neutrino 
 flux from unresolved GRBs labelled  
 in Fig.~\ref{fondo} as GRB pN (1) and (2).
The overall rate of events expected for $E_{\mu}>10$ 
 GeV  is 190 and 46 events per 
  km$^2$ from the   
entire lower Earth hemisphere for the flux (1) and (2) respectively. 
 The event rate for  $E_{\mu}>10$ TeV, in a range where the 
 signal  
 is expected to start arising over the atmospheric neutrino background, 
 becomes 26 and 6 events per km$^2$ 
 respectively. In any case a good knowledge of the background 
 is required in order to have a reliable measure.     

As shown in Fig.~\ref{fondo} and Fig.~\ref{limitiplot}, current experiments  
 have not the required sensitivity to  
 exclude or confirm our predictions. 
 Neutrinos telescopes of the next generation  
 like ANTARES \cite{Antares}, NEMO \cite{nemo}, NESTOR \cite{nestor}, with 
 effective area 
  ranging from $ 0.1$ to 1 km$^{2}$, 
 will be likely capable to success in this task.

\section{Conclusions}

We have presented a model for the production of high energy 
neutrinos from GRBs in an energy range from few GeV up to few PeV,   
 through the interaction of  
 the shock accelerated protons with the matter surrounding the source. 
   The results have been compared with other  
  theoretical predictions  and with the 
  available experimental upper limits from 
   underground and underwater/ice neutrino detectors.   
  Current experiments  
 have not the required sensitivity to  
 exclude or confirm our predictions but  
 the next generation neutrinos telescopes  
 are expected to be capable to reach this goal in few years 
  of data taking.

\section{Acknowledgment} 
We wish to acknowledge the members of the MACRO collaboration, 
 in particular the Lecce group,   
 for the fruitful discussion and the concrete help. 
 A further acknowledgment goes to INFN 
  which contributed to support this research.       
We also thank Prof. P. Galeotti for interesting discussions.







\end{document}